\documentclass[epj]{svjour}
\usepackage{amsmath}
\usepackage{amssymb}
\usepackage{graphics}
\begin{document}
\title{Ferromagnetism in the Periodic Anderson Model}
\subtitle{A Comparison of Spectral Density Approximation (SDA), Modified
  Alloy Analogy (MAA) and Modified Perturbation Theory (MPT)}

\author{D.\ Meyer \and W.\ Nolting}

\institute{Institut f\"ur Physik, Humboldt-Universit\"at zu Berlin,
  Germany}
\date{Received: \today}
%
\abstract{
We compare different approximation schemes for investigating
ferromagnetism in the periodic Anderson
model. The use of several approximations allows for a detailed analysis
of the implications of the respective methods, and also of
the mechanisms driving the ferromagnetic transition.
For the Kondo limit, our results
confirm a previously proposed mechanism leading to ferromagnetic order,
namely an RKKY exchange mediated via the formation of Kondo screening
clouds in the conduction band. The contrary case is found in
the intermediate-valence regime. Here, the bandshift correction ensuring
a correct high-energy expansion of the self-energy is
essential. Inclusion of damping effects reduces stability of the
ferromagnetic phase. 
\PACS{
      {71.10.Fd}{Lattice fermion models (Hubbard model, etc.)}   \and
      {71.28.+d}{Narrow-band systems; intermediate-valence solids} \and
      {75.30.Mb}{Valence fluctuation, Kondo lattice, and heavy-fermion}
     }
} 
\maketitle

\section{Introduction}
\label{intro}
The periodic Anderson model (PAM) provides probably the simplest starting
point to investigate the interplay between 
the hybridization of a periodic array of localized ($f$-) electron levels and
a band of itinerant electrons, the strong correlation of the localized
electrons and quantum mechanics (Pauli principle)\cite{hewson}.

Within this model, we will investigate the many-body problem of ferromagnetic
ordering. Ferromagnetism
in the PAM has previously been examined, but most efforts were aimed
solely at the so-called Kondo regime of the model. This is defined by an
integer number of $f$-electrons per lattice site. These localized electrons
can form an array of well-defined local moments. Via an RKKY
interaction, these can order ferro- or
antiferromagnetically\cite{Jar95,TJF97,MW93,DS98}.

In previous papers, we have shown that in the intermediate-valence
regime of the PAM, defined by a non-integer filling of the localized
levels accompanied by a larger effective hybridization, ferromagnetic
order is also possible\cite{MNRR98,MN99a,RMSRN00pre,mn00c}. This,
however, raises some
questions concerning the driving mechanism for the ferromagnetic
ordering. First of all, in the intermediate-valence regime, quantum
fluctuations effectively delocalize the $f$-electrons, thus the formation
of stable moments needs further explanation. A second question is due
to the fact, that some of the methods used in the above-mentioned papers,
exclude by construction such effects as RKKY as we will discuss
below. So why can these methods give meaningful results when
neglecting something considered as essential as RKKY exchange?
What should be the driving force towards ferromagnetic ordering in
the intermediate valence regime?
In this paper, we try to clarify the apparent incoherence of the
investigations of ferromagnetism in the PAM in the Kondo and the
intermediate-valence regime.

After introducing the model in section~\ref{sec:pam}, we will discuss
several
well-known properties of the system in section~\ref{sec:genprop} and
finally, in 
section~\ref{sec:approx}, introduce a number of approximation schemes. By
comparing
to the known properties discussed before,
the advantages and disadvantages of these methods will become
clear.
In section~\ref{sec:results}, we present and compare
results obtained 
with the different methods. Knowing the strengths and shortcomings of
them will help us to understand the origin of
ferromagnetic order. Altogether, this will lead us to the conclusion
that there are indeed two distinct mechanisms at work: in the Kondo
regime, an RKKY exchange has to be seen as the cause of the ferromagnetic
order. In the intermediate-valence regime, however, the situation
resembles more that of a band-ferromagnet as described by the
single-band Hubbard model. In section~\ref{sec:sum}, we will conclude with a
summary of our findings.

\section{Theory}
\label{sec:theo}

\subsection{The Periodic Anderson Model}
 \label{sec:pam}
We investigate the standard form of the periodic Anderson model (PAM) where
a non-degenerate localized $f$-level hybridizes with a non-degenerate
conduction band (``$s$-band'') via an on-site hybridization:
\begin{align}
  \label{hamiltonian}
    H =&\sum_{\vec{k},\sigma}
    (\epsilon(\vec{k})-\mu)s_{\vec{k}\sigma}^{\dagger}s_{\vec{k}\sigma} + 
    \sum_{i,\sigma} (e_{\rm f}-\mu) f_{i\sigma}^{\dagger}f_{i\sigma} +\\ 
    &V \sum_{i,\sigma} (f_{i\sigma}^{\dagger}s_{i\sigma} +
    s_{i\sigma}^{\dagger}f_{i\sigma} ) + \frac{1}{2} U \sum_{i,\sigma}
    n_{i\sigma}^{(f)}n_{i-\sigma}^{(f)}\nonumber
\end{align}
Here, $s_{\vec{k}\sigma}$ ($f_{i\sigma}$) and
$s_{\vec{k}\sigma}^{\dagger}$ ($f_{i\sigma}^{\dagger}$) are the
creation and annihilation operators for a conduction electron with
Bloch vector $\vec{k}$ and spin $\sigma$ (a localized electron on site
$i$ and spin $\sigma$) and
$n_{i\sigma}^{(f)}=f_{i\sigma}^{\dagger}f_{i\sigma}$.
$s_{\vec{k}\sigma}=\frac{1}{N}\sum_{\vec{k}} e^{i \vec{k} \vec{R}_i}
s_{i\sigma}$ and
$\epsilon(\vec{k})$ is the dispersion of the 
conduction band and $e_{\rm f}$ is the position of the localized level. 
The hybridization strength $V$ is taken to be
$\vec{k}$-independent, and finally, $U$ is the on-site Coulomb
interaction strength between two $f$-electrons. Throughout this paper, the
conduction band will be described by a Bloch (free) density of states,
$\rho_0(E)= \frac{1}{N}\sum_{\vec{k}} \delta (E-\epsilon(\vec{k}))$,
of semi-elliptic shape. Its width $W=1$ sets the
energy scale, and its center of gravity the energy-zero: 
$T_{ii}=\frac{1}{N}\sum_{\vec{k}}\epsilon(\vec k)\stackrel{!}{=}0$.

The problem of determining the relevant (Zubarev) Green functions\cite{Zub60,NolBd7},
\begin{eqnarray}
  \label{eq:greenfunction}
  G_{ij\sigma}^{(f)}(E) = \langle\!\langle f_{i\sigma} ;
  f_{j\sigma}^{\dagger} \rangle\!\rangle;\quad  G_{ij\sigma}^{(s)}(E) =
  \langle\!\langle s_{i\sigma} ; 
  s_{j\sigma}^{\dagger} \rangle\!\rangle\\
G_{\vec k \sigma}^{(f,s)}=\frac{1}{N}\sum_{\vec{k}}
e^{i\vec{k}\cdot(\vec{R}_i-\vec{R}_j)} G_{ij\sigma}^{(f,s)}(E)\label{ggggg}
\end{eqnarray}
can be reduced to the \textit{a priori} unknown self-energy via the
formal solultion of the respective equations of motion:
\begin{subequations}
\begin{align}
  \label{eq:greenfunction2s}
  G_{\vec{k}\sigma}^{(s)}(E) &=
  \frac{E-(e_{\rm f}-\mu)-\Sigma_{\vec{k}\sigma}(E)}{
    (E-(e_{\rm f}-\mu)-\Sigma_{\vec{k}\sigma}(E) ) (E-(\epsilon(\vec{k})-\mu))
    -V^2}\\
  \label{eq:greenfunction2f}
    G_{\vec{k}\sigma}^{(f)}(E) &= 
    \frac{1}
    {E-(e_{\rm f}-\mu)- \frac{V^2}{E-(\epsilon(\vec{k})-\mu)}
      -\Sigma_{\vec{k}\sigma}(E)}  
\end{align}
\end{subequations}
Here, the self-energy is defined by
\begin{equation}
  \label{eq:sigma}
  \Sigma_{\vec{k}\sigma}(E) G_{\vec{k}\sigma}^{(f)}(E) = U \frac{1}{N}
  \sum_{\vec{p},\vec{q}}\langle\!\langle f_{\vec{p}-\sigma}^{\dagger}
  f_{\vec{q}-\sigma} f_{\vec{p}+\vec{k}-\vec{q}\sigma};
  f_{\vec{k}\sigma}^{\dagger} \rangle\!\rangle
\end{equation}
Throughout this paper, we will apply the local approximation, i.\
e.\ assume a $\vec{k}$-independent self-energy. Although
becoming exact only in the limit of infinite spatial dimensions, it 
was shown that this
approximation gives satisfactory results already for three
dimensions\cite{SC89b,SC90a}.

From the Green functions~(\ref{eq:greenfunction2s})
and~(\ref{eq:greenfunction2f}), the $f$- and $s$-quasiparticle densities
of states ($f$- and $s$-DOS) can be calculated:
\begin{subequations}
\begin{eqnarray}
  \label{rhos}
  \rho_{\sigma}^{(s)}(E)=- \frac{1}{\pi N} \sum_{\vec{k}} \Im
  G_{\vec{k}\sigma}^{(s)}(E-\mu+{\rm i} 0^+)\\
  \label{rhof}
  \rho_{\sigma}^{(f)}(E)=- \frac{1}{\pi N} \sum_{\vec{k}} \Im
  G_{\vec{k}\sigma}^{(f)}(E-\mu+{\rm i} 0^+)
\end{eqnarray}
\end{subequations}
The spin-dependent average occupation number
$n^{(s,f)}_{\sigma}$ can now easily be determined:
\begin{subequations}
\label{eq:nf_ns}
\begin{eqnarray}
  \label{ns}
  n_{\sigma}^{(s)}=\langle s_{i\sigma}^{\dagger}s_{i\sigma}\rangle =
  \int_{-\infty}^{+\infty} dE f_-(E) \rho_{\sigma}^{(s)}(E)\\
  \label{nf}
  n_{\sigma}^{(f)}=\langle f_{i\sigma}^{\dagger}f_{i\sigma}\rangle =
  \int_{-\infty}^{+\infty} dE f_-(E) \rho_{\sigma}^{(f)}(E)
\end{eqnarray}
\end{subequations}
Here, $f_-(E)$ denotes the Fermi function and $\Im x$ the imaginary part
of $x$.

Before introducing our approximative methods, let us discuss some
general properties of the PAM.

\subsection{General Properties }
\label{sec:genprop}

\subsubsection{The Hybridization-Free Case}
\label{sec:v=0-case}
For vanishing hybridization strength $V$ (``atomic limit''), the problem
reduces to that of 
the zero-bandwidth Hubbard model\cite{Hub63,NolBd7}. One obtains for the
$f$-Greenfunction:
\begin{equation}
  \label{eq:gf-at}
  G_{\sigma}^{(f,\text{at.})}=\frac{E-(e_{\rm f}-\mu)-U(1-n_{-\sigma}^{(f)})}
    {(E-(e_{\rm f}-\mu))(E-(e_{\rm f}-\mu)-U)}
\end{equation}
with the respective self-energy:
\begin{equation}
  \label{eq:sigma_at}
  \Sigma_{\sigma}^{\text{(at.)}}(E)=\frac{U n_{-\sigma}^{(f)}
    (E-(e_{\rm f}-\mu))}{E-(e_{\rm f}-\mu)-U(1-n_{-\sigma}^{(f)})} 
\end{equation}
The excitation spectrum consists of two peaks located at $e_{\rm f}$ and
$e_{\rm f}+U$, which are called \textit{charge excitations}. The conduction
band DOS remains unchanged.

\subsubsection{The Non-Interacting Limit}
\label{sec:inter-free-limit}
The second trivial limit, the interaction-free limit ($U=0$), yields the
following $f$-Green function:
\begin{equation}
  \label{eq:greenfunctionU0}
    G_{\vec{k}\sigma}^{(f,\text{U=0})}(E) = 
    \frac{1}
    {E-(e_{\rm f}-\mu)- \frac{V^2}{E-(\epsilon(\vec{k})-\mu)}}
\end{equation}
The DOS now consists of two features, one corresponding to the
$f$-level, which becomes broadened due to the hybridization. The other
feature is the renormalized conduction band.
Due to the hybridization there is also an admixture of $f$-spectral weight
into the conduction band region and vice versa.
The amount of this admixture of $f$- and
$s$-states can be understood as a rough estimate of the effective
hybridization. It is
generally stronger for $e_{\rm f}$ close to, or inside the conduction band as
for $e_{\rm f}$ well below the band.
A further effect is clearly visible when
$e_{\rm f}$ lies within the conduction band. Level-repulsion between
$f$-level and the conduction band induces a gap located approximately at
$e_{\rm f}$. 
For $e_{\rm f}$ below the
conduction band,
the level repulsion appears only in form of a small shift of the lower
edge of the conduction band and the center of the $f$-level is situated
slightly below $e_{\rm f}$.

\begin{figure}
  \begin{center}
    \resizebox{0.35\textwidth}{!}{%
      \includegraphics{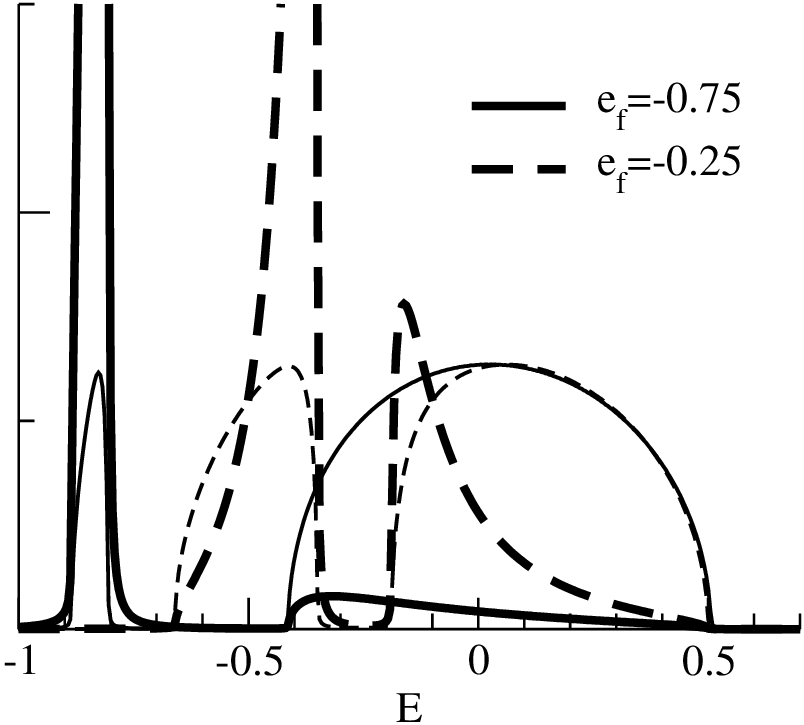}
      }
  \end{center}
\caption{Density of states for $V=0.2$, $e_{\rm f}=-0.75$
  and $-0.25$ in the 
  interaction-free limit ($U=0$). Thick lines: $f$-contribution, thin
  lines conduction band contribution.}
\label{fig:dos_u0}
\end{figure}

These two situations are clearly distinct, one expects different
findings in each. The situation of $e_{\rm f}$ well below the conduction
band leads for finite (large) $U$ to the Kondo regime of the PAM. Here
the $f$-level is almost
integer-filled.
The opposite case, with 
$e_{\rm f}$ located within the band and a hybridization strong enough to
lead to non-integer $n^{(f)}$,
is called 
intermediate-valence regime.
Both situations are plotted in
figure~\ref{fig:dos_u0}. 

\subsubsection{High-Energy Expansion}
\label{sec:high-energy-expans}
Next, we introduce a useful high-energy expansion for the $f$-electron
Green function and the self-energy:
\begin{equation}
  \label{eq:expansion}
  \begin{split}
    G_{\vec{k}\sigma}^{(f)}(E)&=\int dE' \frac{S_{\vec{k}\sigma}(E')} {E-E'}
  = \sum_{n=0}^{\infty} \frac{M_{\vec{k}\sigma}^{(n)}}{E^{n+1}}\\
  \Sigma_{\vec{k}\sigma}(E)&=\sum_{n=0}^{\infty}
  \frac{C_{\vec{k}\sigma}^{(n)}} {E^n}
\end{split}
\end{equation}
$S_{\vec{k}\sigma}(E)=-\frac{1}{\pi} \Im  G_{\vec{k}\sigma}^{(f)}(E)$ is
the spectral density. Its moments, which are the coefficients of the
high-energy expansion of $G_{\vec{k}\sigma}^{(f)}(E)$, are defined by
\begin{equation}
  M_{\vec{k}\sigma}^{(n)}=\int dE E^n
  S_{\vec{k}\sigma}(E);\quad n=0,1,2,\dots 
\end{equation}
and can be calculated independently via
\begin{equation}
  \label{eq:moments}
  M_{\vec{k}\sigma}^{(n)}=\langle  [
  \underbrace{[\dots[f_{\vec{k}\sigma},H]_-,\dots,H]_-}_{\text{$n$-fold
      commutator}}, 
  f_{\vec{k}\sigma}^{\dagger} ]_+\rangle 
\end{equation}
where $[\dots,\dots]_-$ denotes the commutator and $[\dots,\dots]_+$ the
anti-commutator. By inserting (\ref{eq:expansion}) into
(\ref{eq:greenfunction2f}), one can determine the coefficients of the
self-energy expansion. 

For the approximative approaches presented below, 
we only need the local moments 
$M_{\sigma}^{(n)}=\frac{1}{N}\sum_{\vec{k}} M_{\vec{k}\sigma}^{(n)}$ and
coefficients
$C_{\sigma}^{(n)}=\frac{1}{N}\sum_{\vec{k}} C_{\vec{k}\sigma}^{(n)}$:
\begin{subequations}\label{eq:moments2}
  \begin{align}
    \label{eq:moments2_0}
    M_{\sigma}^{(0)} =& 1 \\
    \label{eq:moments2_1}
    M_{\sigma}^{(1)} =& e_{\rm f} + U n_{-\sigma}^{(f)}\\
    \label{eq:moments2_2}
    M_{\sigma}^{(2)} =& e_{\rm f}^2 + 2 e_{\rm f} U n_{-\sigma}^{(f)} +U^2
    n_{-\sigma}^{(f)} +V^2\\
    \begin{split}
    \label{eq:moments2_3}
    M_{\sigma}^{(3)}  =& e_{\rm f}^3 + 3 e_{\rm f}^2 U
      n_{-\sigma}^{(f)} + U^2 e_{\rm f} (2 
      n_{-\sigma}^{(f)} + n_{-\sigma}^{(f)^2}) + U^3
      n_{-\sigma}^{(f)}+\\
      &+ V^2(2 e_{\rm f} +2 U n_{-\sigma}^{(f)} + T_{ii}) +U^2
      n_{-\sigma}^{(f)}  (1-n_{-\sigma}^{(f)}) B_{-\sigma} 
    \end{split}
  \end{align}
\end{subequations}
and
\begin{subequations}
  \label{S_coeff}
  \begin{align}
    \label{S_coeff_0}
    C_{\sigma}^{(0)} &= U n_{-\sigma}^{(f)}\\
    \label{S_coeff_1}
    C_{\sigma}^{(1)} &= U^2 n_{-\sigma}^{(f)} (1-n_{-\sigma}^{(f)})\\
    \label{S_coeff_2}
    C_{\sigma}^{(2)} &= U^2 n_{-\sigma}^{(f)} (1-n_{-\sigma}^{(f)})
    (B_{-\sigma} +U(1-n_{-\sigma}^{(f)}))
  \end{align}
\end{subequations}
The abbreviation $B_{-\sigma}$ in~(\ref{eq:moments2})
and~(\ref{S_coeff}) stands for a higher correlation 
function called \textit{bandshift}:
\begin{equation}
  \label{eq:bandshift}
  n_{-\sigma}^{(f)}(1-n_{-\sigma}^{(f)})(B_{-\sigma}-e_{\rm f}) = V \langle
  f_{i-\sigma}^{\dagger} s_{i-\sigma}(2n_{i\sigma}^{(f)}-1)\rangle
\end{equation}
In spite of the fact that it is a
"higher" correlation function it can rigorously be expressed by
the Green function~(\ref{eq:greenfunction2f}) and the
self-energy~(\ref{eq:sigma})\cite{GN88}:
\begin{gather}
  n_{-\sigma}^{(f)}(1-n_{-\sigma}^{(f)})(B_{-\sigma}-e_{\rm f}) =
  \nonumber \\
  \label{eq:bandshift2}
  - \frac{1}{\pi} \Im \int_{-\infty}^{+\infty} dE f_-(E)
  \left(\frac{2}{U} \Sigma_{\sigma}(E) -1\right)\\
  \times \left(\left(E -(e_{\rm f}-\mu)-
  \Sigma_{\sigma}(E)\right) G_{ii\sigma}^{(f)}(E) -1\right)
\nonumber 
\end{gather}
Surprisingly the hybridization $V$ does not explicitely appear in the
$C_{\sigma}^{(n)}$. The contributions via the moments
(\ref{eq:moments2}) are exactly cancelled by those from the term
$\frac{V^2}{E-(\epsilon(\vec{k})-\mu)}$ in (\ref{eq:greenfunction2f}).
From this, one conclusion can already be drawn: The
hybridization $V$ enters the calculation only via
equation~(\ref{eq:greenfunction2f}) in combination with
the free conduction band dispersion $\epsilon(\vec{k})$. Although there
are clearly correlation-induced effects in the conduction band (cf.\
equation~(\ref{eq:greenfunction2s})), these do not feed back into the
determination of the high-energy features of the self-energy. 
Any RKKY-like indirect exchange between $f$-sites driven by
correlations (cf.\ reference \cite{TJF97}), leaves no footprints in the
high-energy behaviour of the self-energy. Although this reasoning works
only in the local approximation, it does not imply that the local
approximation itself suppresses any RKKY exchange (cf.\ discussion
and references in section~\ref{sec:mpt}).

The prominent high-energy features of the PAM are the charge excitations
known from the zero-hybridization limit. One can easily check that in
this limit the moments~(\ref{eq:moments2}) and self-energy
coefficients~(\ref{S_coeff}) are fulfilled since $B_{-\sigma}\rightarrow
e_{\rm f}$. So obviously, the bandshift, and therefore the $n=3$-moment take
care of a correction of the positions and weights of the charge
excitations in the case of finite interaction and
hybridization. We will show below that this correction can be decisive
for a proper description of ferromagnetism in the PAM.

\subsubsection{Low-Energy Properties}
\label{sec:low-energy-prop}
The PAM is the lattice-periodic extension of the single-impurity
Anderson model (SIAM). Since the latter is famous for its special
low-energy features (``Kondo-physics'')\cite{hewson}, one expects
similar findings also for the PAM.
The most prominent finding for the SIAM is the Kondo-screening: At low
temperatures, the magnetic moment of the impurity-site is screened by
conduction electrons. The remaining conduction electrons form a Fermi
liquid. All physical quantities can be scaled by a single energy:
the Kondo temperature $T_{\rm K}$. In the excitation spectra, the most
significant signature of the screening is the occurence of a sharp
resonance at the chemical potential, the \textit{Kondo Resonance}.

\begin{figure}
  \begin{center}
    \resizebox{0.45\textwidth}{!}{
      \includegraphics{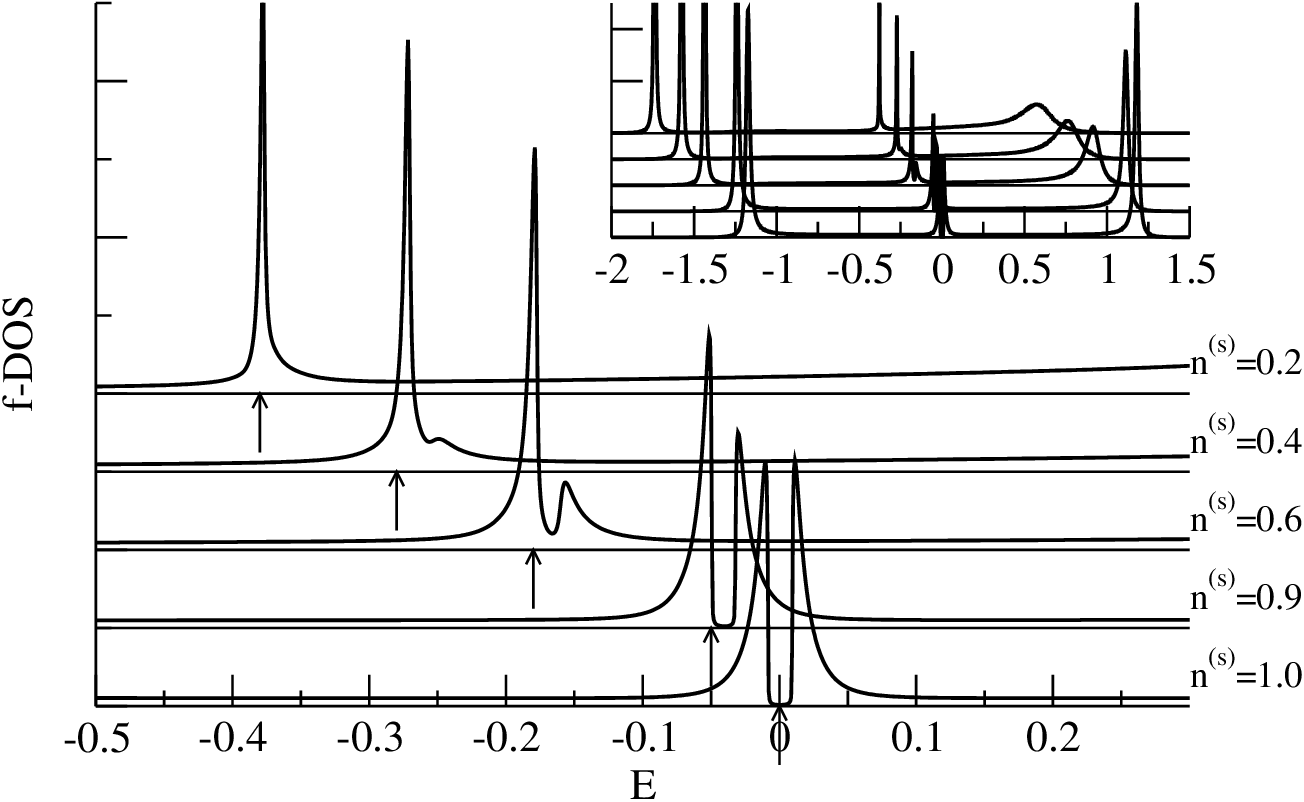}
      }
  \end{center}
\caption{$f$-Density of states as obtained within the MPT (see text)
  for $V=0.2$, $U=2$, $n^{(f)}=1.0$ and
  different $n^{(s)}$ as indicated. $e_f$ was determined such that
  $n^{(f)}=1$ holds.
  The chemical potential is positioned
  at the arrows.
  The inset shows the full energy range, the main picture just the
  region around $\mu$.}
\label{fig:dos_para_nc}
\end{figure}

The low-energy properties of the PAM have been the subject of intense
research\cite{hewson,Fye90,GS91,Jar95,PBJ00b,MN00b}. For the
symmetric PAM, defined
by $n^{(f)}=n^{(s)}=1.0$ and $e_{\rm f}=-\frac{U}{2}$, a Kondo resonance
appears centered at $\mu$. But contrary to the SIAM, it is split by a gap,
the \textit{coherence gap} and the system is insulating. Picturing the
Kondo resonance as virtual $f$-level\cite{And61}, the 
gap is simply due to level-repulsion between the ``flat band'' of the
virtual $f$-levels at every lattice-site and the conduction
band. Although it originates from the same mechanism as the gap discussed
in section~\ref{sec:inter-free-limit}, it is clearly distinguishable:
The coherence gap is, together with the Kondo resonance, pinned at $\mu$ whereas the
hybridization gap discussed in the interaction-free limit would show up
at $e_{\rm f}$. 
In figure~\ref{fig:dos_para_nc}, we present the densities of states as
obtained with the below-discussed modified perturbation theory (MPT).
Although these DOS are calculated within
an approximative method, the qualitatively same picture emerges from the 
numerically exact
Quantum Monte Carlo (QMC)\cite{Jar95,TJF98} and numerical
renormalization group calculations (NRG)\cite{PBJ00b} and can thus be
believed to be qualitatively correct.

The situation of the
symmetric PAM is plotted at the bottom of figure~\ref{fig:dos_para_nc}.
Moving away from the symmetric PAM, e.g.\  by reducing the number of
conduction electrons, the system becomes
metallic again. The Kondo resonance gets asymmetric relative to
$\mu$, the coherence gap is shifted away from the chemical potential.
Further away from the symmetric parameters, the coherence gap closes,
which is probably due to quasiparticle damping, since $\Im
\Sigma_{\sigma}(E) \sim E^2$.

A still open question is that of a unique energy scale similar to
$T_{\rm K}$ for the SIAM. The main problem under discussion is the
\textit{exhaustion problem} introduced by Nozi{\`e}res\cite{Noz98}. For
the case of a periodic array of localized $f$-levels, the system cannot
make available enough conduction electrons to screen all
$f$-moments. This situation is still subject to many
investigations\cite{TJF97,TJF98,PBJ00b,MN00b,BGG00pre}. 

\subsection{Approximative Solution of the PAM}
\label{sec:approx}
In the following sections, we will introduce a series of approximation
methods to determine the self-energy~(\ref{eq:sigma}) of the periodic
Anderson model. This series represents a subsequent improvement of the
theory along the lines discussed in the last section: The Hubbard-I
approximation recovers the non-interacting and the hybridization-free
limit but fails to reproduce the high-energy behaviour of the
self-energy. This is corrected by the spectral density
approximation. This method, however, still suffers from the complete
neglection of quasiparticle damping. A straightforward method to
incorporate this is the modified alloy analogy. Finally, the low-energy
behaviour can be qualitatively reproduced by the modified perturbation
theory.

\subsubsection{The Hubbard-I Approximation}
\label{sec:HI}
The first approach presented here is a crude interpolation between the
two exactly solvable limiting cases $V=0$ (cf.\
section~\ref{sec:v=0-case}) and $U=0$ (cf.\
section~\ref{sec:inter-free-limit}). The result equals that of Hubbard's
first work on the Hubbard-model\cite{Hub63}, the Hubbard-I approximation.

In the interaction-free limit, one can express the $f$-Green
function~(\ref{eq:greenfunctionU0}) in terms of the corresponding
$U=0$-$V=0$ (``atomic'') solution:
\begin{equation}
  \label{eq:gf0-at}
    G_{\vec{k}\sigma}^{(f,\text{U=0})}(E) = 
    \frac{1}
    {\left(G_{\sigma}^{(f,\text{U=0,at.})}(E)\right)^{-1}-
      \frac{V^2}{E-(\epsilon(\vec{k})-\mu)}} 
\end{equation}
The Hubbard-I approximation is now obtained by assuming the functional
dependence of (\ref{eq:gf0-at}) also for the finite-$U$ case.
With the atomic-limit Green function for the full PAM~(\ref{eq:gf-at}),
this essentially corresponds to inserting the
self-energy~(\ref{eq:sigma_at}) into
equation~(\ref{eq:greenfunction2f}), therefore
\begin{equation}
  \label{eq:sigma_HI}
    \Sigma_{\sigma}^{\text{(H-I)}}(E)=\Sigma_{\sigma}^{\text{(at.)}}(E)
    =\frac{U n_{-\sigma}^{(f)}
      (E-(e_{\rm f}-\mu))}{E-(e_{\rm f}-\mu)-U(1-n_{-\sigma}^{(f)})} 
\end{equation}

Although this method is by construction exact in two limiting cases,
namely the interaction-free ($U=0$) and the hybridization-free ($V=0$)
case, a number of shortcomings follows directly from inspecting
equation~(\ref{eq:sigma_HI}) and its derivation. First of all, the
self-energy~(\ref{eq:sigma_HI}) fulfills the high-energy expansion only
up to the $n=1$ self-energy coefficient. The bandshift correction
$B_{-\sigma}$ is neglected.  This yields a weak
spin-dependence through the expectation values 
$n_{-\sigma}^{(f)}=\langle n_{i-\sigma}^{(f)}\rangle$ only.
Second, the self-energy~(\ref{eq:sigma_HI})
is real, any quasiparticle damping effects as indicated by a
finite imaginary part of $\Sigma_{\sigma}(E)$ are ignored.
The third drawback is the suppression of any feedback mechanism from the
conduction band:
The latter enters the calculation only via equation~(\ref{eq:gf0-at}). This
implies that although there are correlation-induced changes in the
conduction band (cf.\ equation~(\ref{eq:greenfunction2s})), these do
not feedback into the $f$-self-energy~(\ref{eq:sigma_HI}). An indirect
magnetic exchange between the $f$-electrons via polarization of the
conduction band (``RKKY'') cannot appear in this approximation method.
Finally, none of the expected low-energy features of the PAM, as
discussed in section~\ref{sec:low-energy-prop} can be found
using this method.

\subsubsection{The Spectral Density Approximation}
\label{sec:sda}
The spectral density approximation (SDA)\cite{Nol72,NolBd7,HN97a} is the result of a direct
improvement of the above-discussed Hubbard-I approximation with respect
to the high-energy expansion~(\ref{eq:expansion}).

Starting with an
ansatz for the self-energy using the same functional structure as
(\ref{eq:sigma_HI}),
\begin{equation}
  \label{eq:ansatz_sda}
  \Sigma_{\sigma}(E)= \alpha_{1\sigma} \frac{E-\alpha_{2\sigma}}
{E-\alpha_{3\sigma}}
\end{equation}
one can fit the coefficients $\alpha_{p\sigma}$ 
in such a way that the high energy expansion of
the self-energy ~(\ref{eq:expansion}) with the coefficients
(\ref{S_coeff}) is fulfilled. One readily arrives at 
\begin{equation}
  \label{eq:sigma_sda}
  \Sigma_{\sigma}^{(\rm SDA)}(E)=\frac {U n_{-\sigma}^{(f)}
    (E-B_{-\sigma}-(e_{\rm f}-\mu))} { E-B_{-\sigma} -(e_{\rm f}-\mu) -U
    (1-n_{-\sigma}^{(f)})} 
\end{equation}
The SDA self-energy differs from the Hubbard-I solution by the
bandshift $B_{-\sigma}$ (\ref{eq:bandshift}). It is introduced by the
$n=3$-moment~~(\ref{eq:moments2_3}) and responsible to reproduce the
correct high-energy behaviour of the $f$-Green function.
It leads to a (possibly spin-dependent) shift of
the positions of the $f$-peaks in the density of states. 
So without loosing any of the advantages of the Hubbard-I approximation,
as the correct reproduction of the $U=0$ and the $V=0$ limits and its
numerical simplicity, one major drawback of this method can be removed.
However, the other points of criticism, as the missing quasiparticle
damping and the incorrect low-energy properties remain.

The name ``spectral density approximation'' stems from its application
to the Hubbard model, where this approach is derived by a physically
motivated two-pole ansatz for the spectral
density\cite{Nol72,HN97a}. Results for the SDA in the context of the PAM
were previously published\cite{MNRR98,MN99a}. In these papers, the SDA
was applied to a set of effective Hubbard models, onto which the PAM
could be mapped. However, the above-described procedure leads to exactly
the same results as those published in references~\cite{MNRR98,MN99a}

\subsubsection{The Modified Alloy Analogy}
\label{sec:maa}
Now we want to present a method that resolves one major drawback of the
already introduced methods: Both the Hubbard-I and the SDA
self-energies are real. Quasiparticle damping as represented by a
finite imaginary part of the self-energy is therefore completely
neglected. A well-known method to incorporate damping effects is the
alloy-analogy approach. By using physical intuition or by other
justified means, the original problem is mapped onto an fictitious
alloy, which subsequently can be solved using standard methods as e.\
g.\ the coherent potential approximation (CPA)\cite{VKE68}.

The CPA represents the best ``single-site'' method for solving an
alloy-problem\cite{NolBd7}. Single-site approximation in this context is
equivalent to the already introduced local approximation or
$\vec{k}$-independence of the self-energy.
The alloy is defined by the energy
levels of its components, $E_{p\sigma}$ and their respective concentrations
$x_{p\sigma}$, where the index $p$ numbers the components. The corresponding
self-energy can be determined by solving the CPA-equation:
\begin{equation}
  \label{cpa}
  0=\sum_{p=1}^{n} x_{p\sigma}\frac{E_{p\sigma}-\Sigma_{\sigma}(E)-e_{\rm f}}
  {1-G_{ii\sigma}^{(f)}(E) (E_{p\sigma}-\Sigma_{\sigma}(E)-e_{\rm f})}
\end{equation}
In the conventional alloy analogy for the PAM\cite{LC79b,MA79,Czy86}, the
artificial alloy is determined by the poles of the atomic limit
$f$-Green function~(\ref{eq:gf-at}) and their respective weights:
\begin{align}
  \label{aa}
  E_{1\sigma}^{(AA)}&=e_{\rm f};&   x_{1\sigma}^{(AA)}&=1-n_{-\sigma}^{(f)}\\
  E_{2\sigma}^{(AA)}&=e_{\rm f}+U;& x_{2\sigma}^{(AA)}&=n_{-\sigma}^{(f)}\nonumber 
\end{align}
This choice, however, is in no way predetermined.
In reference~\cite{RMSRN00pre},
another alloy analogy, the modified alloy analogy (MAA) was proposed for
the PAM in
analogy to the MAA for the Hubbard model\cite{HN96}: By
inserting equation~(\ref{eq:expansion})
into~(\ref{cpa}) and comparing the coefficients in $\frac{1}{E}$, an
optimum two-component alloy analogy with respect to the high-energy
behaviour can be found:
\begin{align}
  \label{maa}
  \begin{split}
    E_{1,2\sigma}=&\frac{1}{2} [ B_{-\sigma}+U+e_{\rm f} \pm\\
    &\sqrt{(B_{-\sigma}+U-e_{\rm f})^2+ 4 U n_{-\sigma}^{(f)}
      (e_{\rm f}-B_{-\sigma})}]
  \end{split}\\
  x_{1\sigma}=&\frac{E_{2\sigma}-e_{\rm f}-Un_{-\sigma}^{(f)}}{E_{2\sigma} -
    E_{1\sigma}} = 1-x_{2\sigma}\nonumber
\end{align}
It should be noted that $E_{1,2\sigma}$ coincide with the poles of the
SDA Green function (cf.\ equations~(\ref{eq:sigma_sda})
and~(\ref{eq:greenfunction2f})) when the term
$\frac{V^2}{E-(\epsilon(\vec{k})-\mu)}$ in the denominator of
the Green function is neglected.

By construction, the MAA fulfills the high-energy
expansion~(\ref{eq:expansion}) up to the $n=3$ moment or equivalently the
$n=2$ self-energy coefficient~(\ref{S_coeff_2}). Also, damping effects are
considered. However, the MAA still suffers from the complete neglection
of the low-energy properties (cf.\ section~\ref{sec:low-energy-prop})
and the self-energy is determined in such a way that correlation effects
within the conduction band do have no influence (``feedback'') onto
$\Sigma_{\sigma}(E)$. 
A more detailed discussion of the MAA applied to the PAM can be found
in reference~\cite{RMSRN00pre}.

\subsubsection{The Modified Perturbation Theory}
\label{sec:mpt}
Let us finally introduce the modified perturbation theory. This approach
is based on the dynamical mean-field theory
(DMFT)\cite{PJF95,GKKR96}. In the case of a local self-energy, 
which becomes exact in the limit of
infinite spatial dimensions\cite{MV89,Mue89}, the PAM can be mapped onto
a single-impurity Anderson model (SIAM) with the Hamiltonian,
\begin{align}
  \label{SIAM}
    H =&\sum_{\vec{k},\sigma}
    (\epsilon(\vec{k})-\mu)s_{\vec{k}\sigma}^{\dagger}s_{\vec{k}\sigma} + 
    \sum_{\sigma} (e_{\rm d}-\mu) d_{\sigma}^{\dagger}d_{\sigma} +\\ 
    & \sum_{\sigma}V_{\vec{k}d} (d_{\sigma}^{\dagger}s_{\vec{k}\sigma} +
    s_{\vec{k}\sigma}^{\dagger}d_{\sigma} ) + \frac{1}{2} U \sum_{\sigma}
    n_{\sigma}^{(d)}n_{-\sigma}^{(d)}\nonumber
\end{align}
The notation is as for the
PAM (cf.\ equation~(\ref{hamiltonian})). $d_{\sigma}^{(\dagger)}$ are
the annihilation (creation) operators for electrons at the impurity, its
energy
level is $e_{\rm d}$. In the case of the DMFT, the bath of conduction
electrons, usually defined by $\epsilon(\vec{k})$ and $V_{\vec{k}d}$ need
not be specified in detail. For all practical calculations, it is
sufficient to know the \textit{hybridization function} $\Delta(E)$,
in the pure SIAM defined by $\Delta(E)=\sum_{\vec{k}}
\frac{V_{\vec{k}d}^2}{E-\epsilon(\vec{k})}$. However, for the mapping of
the DMFT to be successfull, the hybridization function has to be
determined according to the \textit{self-consistency condition}\cite{GKKR96}
\begin{equation}
  \label{eq:selfconsistency}
    \Delta_{\sigma}(E)= E-(e_{\rm f}-\mu)-\Sigma_{\sigma}(E)
  -\left(G_{ii\sigma}^{(f)}(E)\right)^{-1} 
\end{equation}
This implies that in case of symmetry breaking, the hybridization
function becomes spin-dependent. Now one can make use of the fact that
the impurity self-energy of the SIAM defined by
equation~(\ref{eq:selfconsistency}) is equivalent to the self-energy of
the PAM.
The advantage of the mapping is that the SIAM is one of the simplest
know many-body models, several exact statements as well as well-tested
approximative solutions are known. In the following, we apply the
modified perturbation theory, which was explained and discussed in detail
elsewhere\cite{PWN97,MWPN99,MN00b,mn00c}. So we will restrict
ourselves to a short summary of this approach.

Starting point is the following ansatz for the
self-energy\cite{MR82,MR86,KK96}:
\begin{equation}
  \label{eq:ansatz}
  \Sigma_{\sigma}(E)=U \langle n_{-\sigma}^{(f)}\rangle
  +\frac{\alpha_{\sigma} \Sigma_{\sigma}^{\rm (SOC)}(E)}
  {1-\beta_{\sigma} \Sigma_{\sigma}^{\rm (SOC)}(E)}
\end{equation}
$\alpha_{\sigma}$ and
$\beta_{\sigma}$ are introduced as parameters to be determined
later. $\Sigma_{\sigma}^{\rm (SOC)}(E)$ is the
second-order contribution to perturbation theory around the Hartree-Fock 
solution\cite{Yam75,ZH83,SC90a}. 
Equation~(\ref{eq:ansatz}) can be understood as the simplest possible
ansatz which can, on the one hand, reproduce the perturbational result
in the limit $U\rightarrow 0$, and, on the other hand, recovers the
atomic limit for
appropriately chosen $\alpha_{\sigma}$ and $\beta_{\sigma}$\cite{LMF99pre}.

Using the perturbation theory around the Hartree-Fock solution
introduces an ambiguity into the calculation. Within the self-consistent
Hartree-Fock calculation, one can either choose the chemical potential
to be equivalent to the chemical potential of the full MPT calculation,
or take it as parameter $\tilde{\mu}$ to be fitted to another physically
motivated constraint.
In reference~\cite{KK96} and other papers~\cite{TS99b,VTJK00}, the
Luttinger theorem\cite{LW60} or equivalently the Friedel sum
rule\cite{Fri56,Lan66} was used to determine $\tilde{\mu}$.
Since these theorems are applicable
only for $T=0$, this limits the calculations to zero temperature. In
order to access finite temperatures, we used the condition of
identical electron densities for the Hartree-Fock and 
the full calculation ($n_{\sigma}^{(f,{\rm HF})}=n_{\sigma}^{(f)}$). In
our view, it is more reasonable to perform the Hartree-Fock calculation
for the same electron density rather than identical chemical potential of
the full MPT calculation since the electron density is a critical
parameter concering correlation effects.
A more detailed analysis of the
different possibilities to determine $\tilde{\mu}$ is found in
reference~\cite{PWN97}. 
Finally, the parameters $\alpha_{\sigma}$ and $\beta_{\sigma}$
have to be determined. Instead of using the ``atomic'' limit of $V=0$ as 
was done e.\ g.\ in references~\cite{KK96,VTJK00,Sas00}, we
make use of the moments of the spectral density. Analogously to
equations~(\ref{eq:expansion}),~(\ref{eq:moments}) and
(\ref{eq:moments2}), these can be evaluated for the SIAM. To fit the two
parameters of ansatz~(\ref{eq:ansatz}), the first three self-energy
coefficients are needed, since $C_{\sigma}^{(0)}$ is reproduced for any
choice of $\alpha_{\sigma}$ and $\beta_{\sigma}$.

As for the PAM, a bandshift correlation function similar
to~(\ref{eq:bandshift}) is introduced via $C_{\sigma}^{(2)}$ and the
procedure leads to
the correct high-energy behaviour of the Green function for the SIAM and
via the DMFT-mapping also for the PAM.
So while recovering the main advantage of the SDA and MAA, namely the
correct reproduction of the high-energy expansion~(\ref{eq:expansion})
up the $n=3$-moment, the MPT yields a major improvement concerning the
low-energy properties of the PAM.
Although already for the SIAM the low-energy scale
(``Kondo temperature'' $T_{\rm K}$) connected with these properties,
cannot be quantitatively reproduced, 
other quantities can, at least in a qualitatively satisfactory way, be
recovered\cite{MWPN99,MN00b}. In particular, the densities of states
both above and below the Kondo temperature, but also the general
features of the susceptibility
$\chi(T)$ seem to be trustworthy.
Another test of the low-energy properties is given by the Friedel
sum rule, which  links the self-energy at the Fermi energy
with the electron density. Within the MPT, it is fulfilled in a large
parameter space\cite{MWPN99}.
It is also worth mentioning that via the DMFT
self-consistency~(\ref{eq:selfconsistency}), a feedback from
correlation-induced features in the conduction band onto the
$f$-self-energy is possible. 
In the limit of infinite spatial dimension ($d\rightarrow\infty$), where
the DMFT becomes
exact, the RKKY exchange between any two lattice sites will
vanish. However, as discussed in reference~\cite{Jar95}, the net
exchange of one lattice site with a shell of neighburs remains finite
since the number of sites in the respective shell diverges as
$d\rightarrow\infty$. This exchange is incorporated in the
(spin-dependent) hybridization function~(\ref{eq:selfconsistency}).
\begin{figure*}
  \begin{center}
    \resizebox{0.7\textwidth}{!}{
      \includegraphics{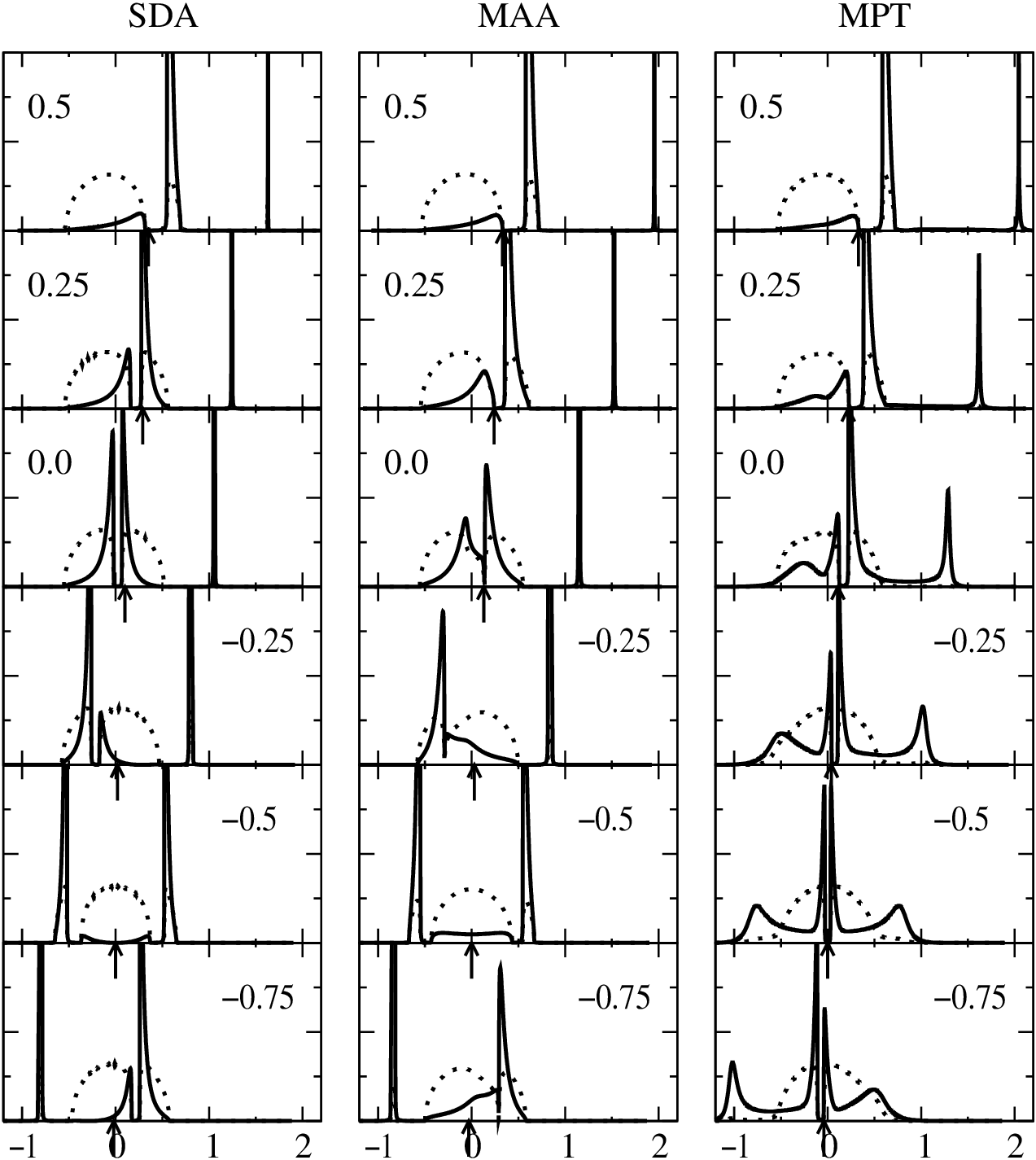}
      }
  \end{center}
\caption{$s$-DOS ($f$-DOS) as dotted (solid) lines for
  $V=0.1$, $U=1$ and $n^{\text{(tot)}}=2.0$. The respective value of
  $e_f$ is given in each graph.
  The position of the chemical potential is denoted by the arrows.
  The left column was calculated with the SDA, the middle within MAA and
  the right column within MPT.}
\label{fig:dos_para}
\end{figure*}

\section{Results and Discussion}
\label{sec:results}
In the following we will present the results obtained with the
different approximation schemes of section~\ref{sec:approx} and hope to
shed some light on the mechanism that leads to ferromagnetism in the PAM
in the Kondo- and intermediate-valence regime. First however, we will
look at the paramagnetic quasiparticle densities of  states (DOS) as
defined by equations~(\ref{rhos}) and~(\ref{rhof}).

In figure~\ref{fig:dos_para}, both the $f$- and $s$-DOS are
plotted for a relatively small interaction strength $U=1.0$ and $V=0.2$,
$n^{\text{(tot)}}=2.0$ at zero temperature and for various $e_{\rm f}$ as
indicated. The second picture from the bottom represents the
above-introduced symmetric case with $e_{\rm f}=-0.5=-\frac{U}{2}$.
The position of the chemical
potential $\mu$ is indicated by the arrows. The left column was obtained
using the SDA, the middle MAA and the right column by using the MPT.

The SDA DOS differs from the interaction-free case (cf.\
section~\ref{sec:inter-free-limit}) plotted in
figure~\ref{fig:dos_u0} by the appearance of a second charge excitation
approximatly at $e_f+U$.
If
either $e_{\rm f}$ or
$e_{\rm f}+U$ falls within the band region, a hybridization gap as discussed
in section~\ref{sec:inter-free-limit} is clearly visible.
The Hubbard-I results are not shown in figure~\ref{fig:dos_para}. These
look very similar to the SDA DOS, only a small shift of the charge
excitations can be noticed. In the symmetic case they are identical,
the bandshift vanishes.

The MAA DOS shows some modifications when compared to the SDA: The
quasiparticle damping softens the charge excitations and the hybridization
gap is for $e_{\rm f}\lesssim 0$ almost closed. For $e_{\rm f}\gtrsim
0.25$ the DOS
strongly resembles the SDA results. This can be understood since the
number of $f$-electrons, $n^{(f)}$ is very small. Scattering processes
become rare and the quasiparticle damping, which differentiates
between SDA and MAA, negligible.
As in the case of the SDA and Hubbard-I approximation, the simpler
theory disrespecting the high-energy expansion and
neglecting the bandshift correction, in this case the conventional alloy
analogy~(\ref{aa}), yields very similar and in the symmetric
case identical results. We will see below that the bandshift correction
becomes much more important in the ferromagnetic
phase. 
\begin{figure}
  \begin{center}
    \resizebox{0.45\textwidth}{!}{
      \includegraphics{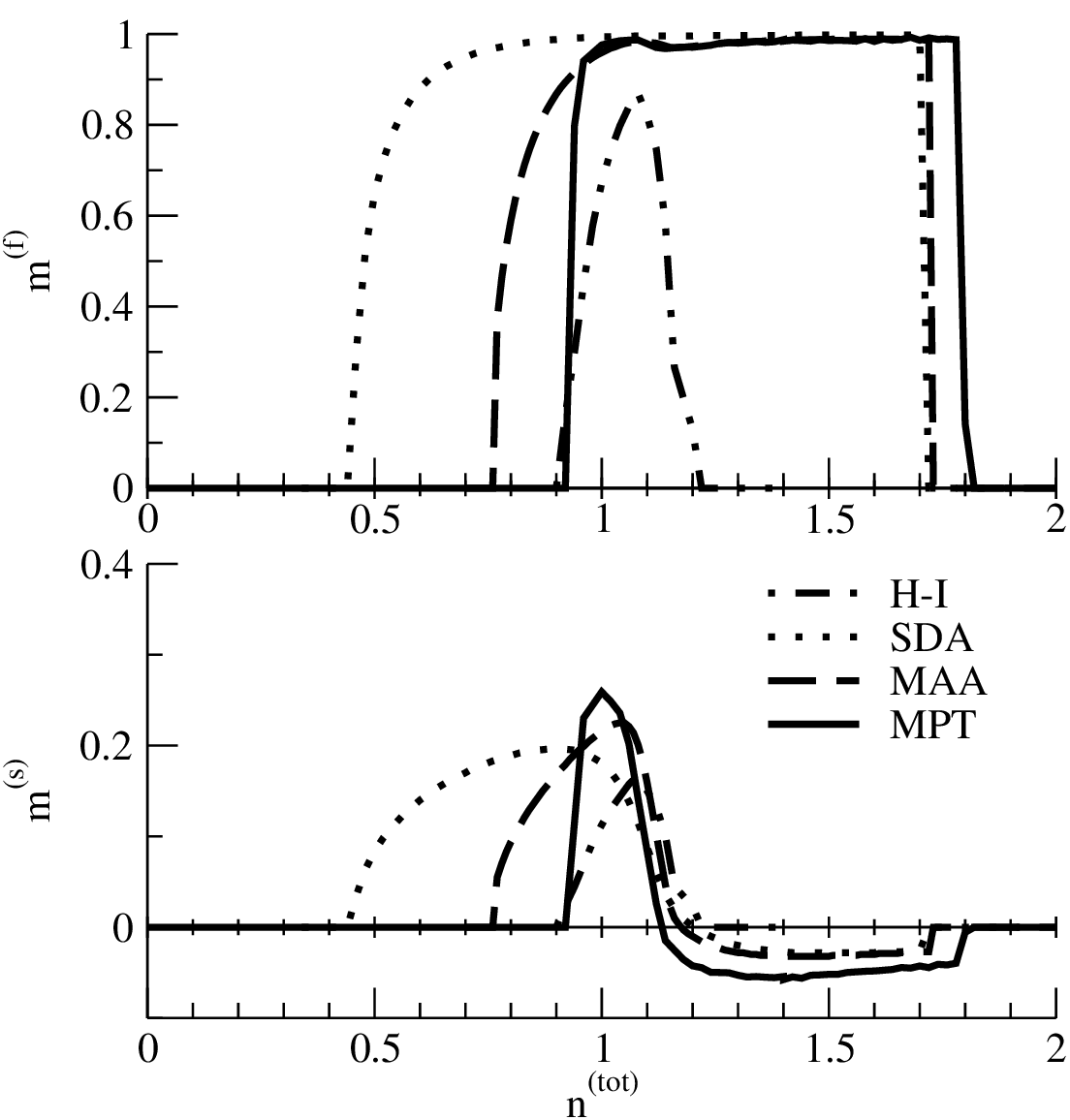}
      }
  \end{center}
\caption{$s$- and $f$- polarization for $U=4$,
  $V=0.1$, $T=0$ and $e_{\rm f}=-0.35$ ($s$: lower, $f$: upper
  panel). The results of Hubbard-I, SDA, MAA and MPT are shown. Please
  note the different scales of the y-axis.}
\label{fig:m_n}
\end{figure}

Finally, the DOS obtained by MPT shows remarkable differences, especially
close to the chemical potential. These represent the ``Kondo physics''
discussed in section~\ref{sec:low-energy-prop}. Again, we note
that for $e_{\rm f}\gtrsim 0.25$ the DOS resemble the SDA
and MAA results. This is obviously for the same reasons as discussed
above, namely the small number of $f$-electrons. 

In figure~\ref{fig:m_n}, the $f$- and $s$-magnetization is plotted as
function of the total electron density $n^{\text{(tot)}}$ for $U=4$,
$e_{\rm f}=-0.35$, $V=0.1$ and $T=0$. Within the Hubbard-I approximation, a
small region of ferromagnetism is found around $n^{\text{(tot)}}\approx 1$. The
conduction band magnetization $m^{(s)}$ (thin lines in
figure~\ref{fig:m_n}) is always positive. We call this situation
parallel $s$-$f$ coupling.

In the SDA the region of ferromagnetism is strongly
enlarged. This is a clear indication of the importance of the bandshift
correction with respect to ferromagnetism. Since the
bandshift~(\ref{eq:bandshift}) can be spin-dependent, it enhances the
possibility of ferromagnetic ordering. We also note an interesting
behaviour of the conduction band magnetization: as function of electron
density, it changes sign. For low $n^{\text{(tot)}}$, the $s$-$f$ coupling is
parallel, for higher values antiparallel. This behaviour,which is also
found within the MAA and MPT, can be traced back to the appearance of
the  hybridization gap discussed in
section~\ref{sec:inter-free-limit}. A more detailed investigation of
this can be found in references~\cite{mn00c,RMSRN00pre}.

\begin{figure}
  \begin{center}
    \resizebox{0.45\textwidth}{!}{
      \includegraphics{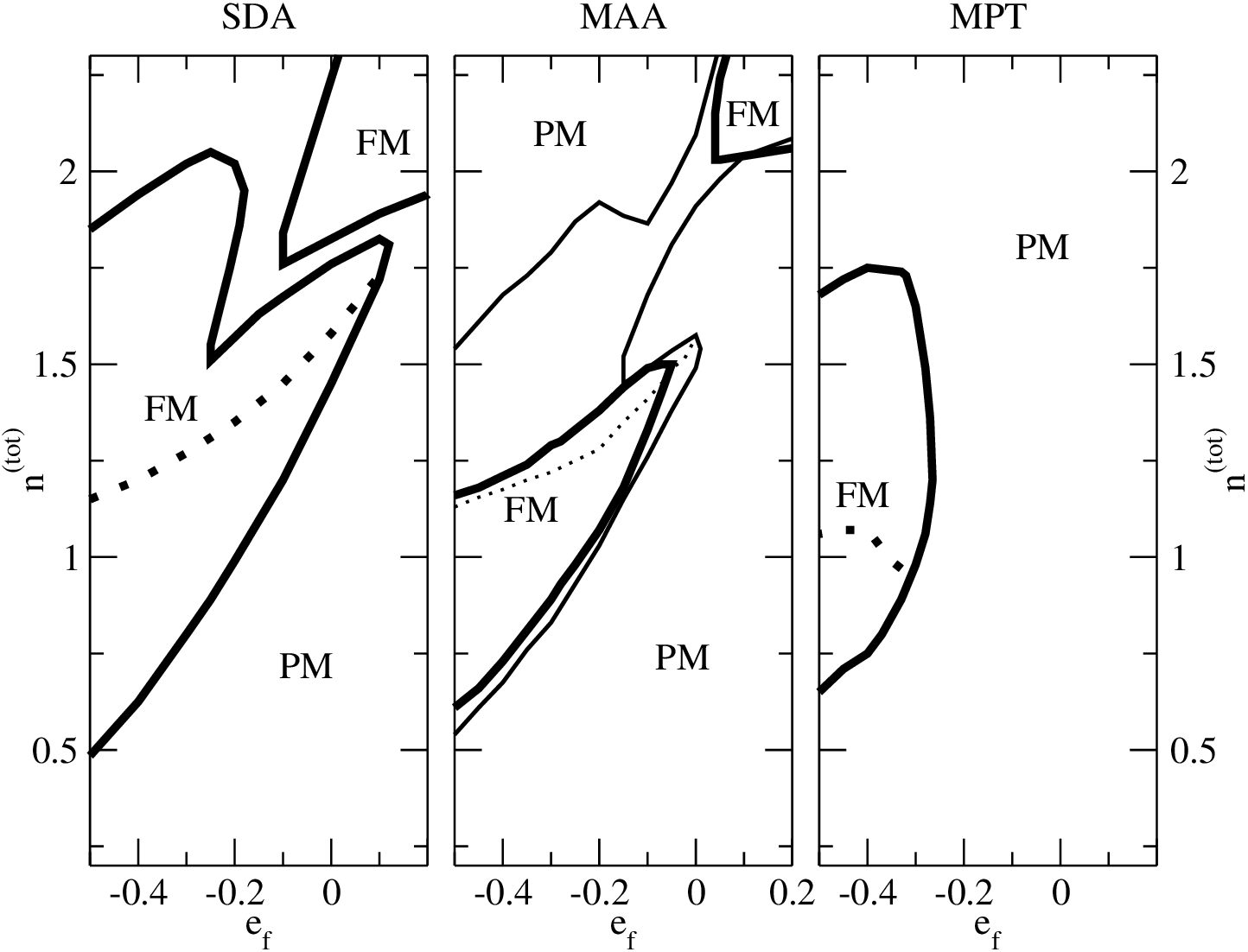}
      }
  \end{center}
\caption{$T=0$ phase diagram for the PAM in the intermediate-valence
  regime with $U=4$ and $V=0.2$ as obtained within SDA, MAA and MPT
  calculations. For the MAA, the phase diagram for $V=0.1$ is
  additionally plotted as thin lines. The dotted lines separate the
  regions of parallel (below the dotted line) and anti-parallel (above)
  $s$-$f$ coupling. Note: for the MAA with $V=0.2$, the low-density
  ferromagnetic region exhibits only parallel, the high-density region
  only antiparallel $s$-$f$-coupling.} 
\label{fig:phdia}
\end{figure}
The MAA result can be used to estimate the influence of quasiparticle
damping. One observes a reduction of the ferromagnetic region compared
to the SDA result. So,
similar to the Hubbard model\cite{HN96,PHWN98}, quasiparticle damping is
unfavourable for ferromagnetism. However, apart from the lower critical
$n^{\text{(tot)}}$, the MAA and SDA curves are very similar.
It should be noted that with the conventional alloy analogy~(\ref{aa}),
no ferromagnetic solution can be found\cite{LC79b}. This confirms again the
importance of the bandshift correction~(\ref{eq:bandshift}) with respect
to ferromagnetism.

The same holds true for the MPT result. Again, the region of
ferromagnetism is reduced as compared to the MAA, but the change
is rather small. Qualitative features as the change of sign of the
conduction band magnetization, and the generally larger conduction band
polarization in case of parallel coupling remain the same for SDA, MAA
and MPT. It should further be pointed out that the lower critical
$n^{\text{(tot)}}$ is, when varying $e_{\rm f}$, in fact determined by
$n^{(f)}$. The number of conduction band electrons, $n^{(s)}$ plays no
significant role\cite{mn00c}.

From all these observations we conclude that the bandshift correction has a
strong influence on
ferromagnetism. So does the inclusion of quasiparticle damping. However,
at least in the examined situation of the PAM in the
intermediate-valence regime, there seems to be no major difference
between the MAA without, and the MPT including the special low-energy
properties of the PAM. These have apparently less influence on
ferromagnetism in the intermediate-valence regime of the PAM.

A quite similar picture emerges from an inspection of the phase
diagram in figure~\ref{fig:phdia}. A comparison of SDA and MAA results
shows the
negative influence of quasiparticle damping on ferromagnetism. Furthermore,
another important point should be noticed here: in the MAA calculation,
a hybridization strength of $V=0.2$ already strongly reduces the magnetic
region compared to the $V=0.1$ case. It is apparent that only
the region with antiparallel $s$-$f$ coupling is affected by this. The
region with parallel $s$-$f$ coupling remains almost unchanged.
It
was shown in reference~\cite{RMSRN00pre} that the MAA shows an anomalous
hybridization dependence in this region. Whereas for other electron
densities, an
increasing hybridization strength $V$ quickly suppresses
ferromagnetism\cite{RMSRN00pre}, here the Curie temperature increases
with increasing $V$ after going through a minimum.
The same behaviour is found within the
SDA. 
However, the MPT does not show this behaviour at all. As discussed in
reference~\cite{MN00b}, the local moments get quenched by formation of
local Kondo singlets with increasing $V$. Being of low-energy nature,
this effect is not covered by the other approximation schemes. The
apparent stability of ferromagnetism around $n^{(tot)}\approx 1$ in these
methods for large hybridization strengths seems therefore to be rather
meaningless. 

A further difference between MAA and MPT is the upper critical $f$-level
position $e_{\rm f}$. The modifications in the MPT are due to the fact that
the lower charge excitation joins with the Kondo resonance. The
existence of the latter is neglected in the MAA. The MPT seems to be
more reliable for determining this phase boundary.

\begin{figure}
  \begin{center}
    \resizebox{0.45\textwidth}{!}{
      \includegraphics{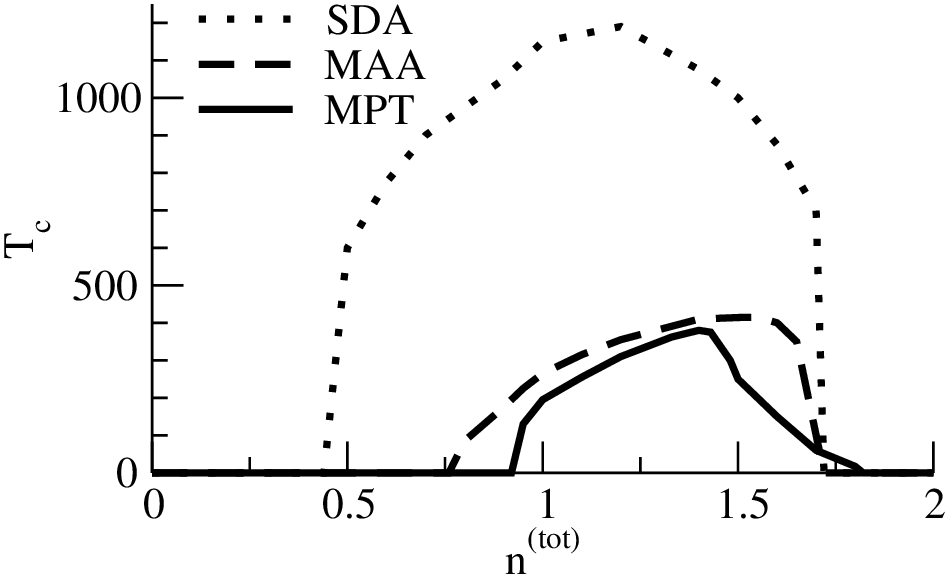}
      }
  \end{center}
\caption{The Curie temperature as function of total electron density for
  $U=4$, $V=0.1$ and $e_{\rm f}=-0.35$.}
\label{fig:tc_n}
\end{figure}
In figure~\ref{fig:tc_n}, the Curie temperatures as function of the
total electron density are plotted for the same model parameters as used
in figure~\ref{fig:m_n}. Again the conclusions are consistent: The
quasiparticle damping which basically discriminates SDA and MAA, leads
to a huge reduction of $T_{\rm c}$. The inclusion of the low-energy
physics, as done by the MPT, does not change $T_{\rm c}$ much, only above
$n^{\text{(tot)}}\approx 1.5$, a suppression of $T_{\rm c}$ is observed.
It is further noteworthy that the change of sign of the conduction band
magnetization as seen in figure~\ref{fig:m_n} does not lead to any
particularity in the $T_{\rm c}$ curves.

\begin{figure}
  \begin{center}
    \resizebox{0.45\textwidth}{!}{
      \includegraphics{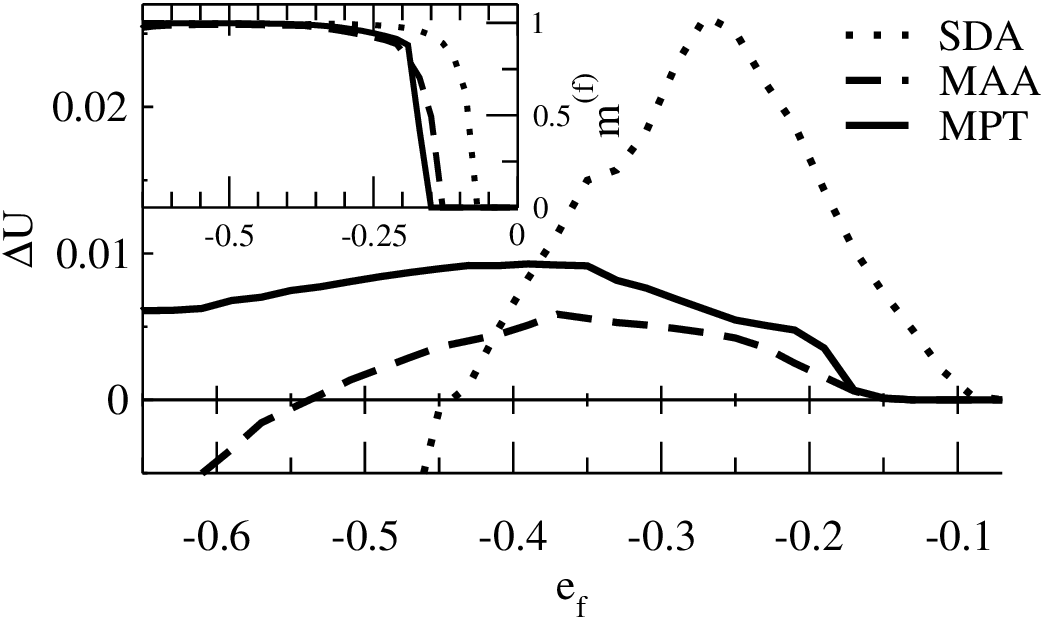}
      }
  \end{center}
\caption{Difference of the internal energy for the paramagnetic and
  ferromagnetic solution (see text). The inset shows the
  $f$-magnetization of the ferromagnetic solution. All calculations for
  $U=4$, $V=0.1$, $T=0$ and $n^{\text{(tot)}}=1.2$.}
\label{fig:m_ef}
\end{figure}

Up to now we have focused on the PAM in the intermediate-valence
regime. What happens to
the ferromagnetic solution upon entering the Kondo regime? The situation
is plotted in figure~\ref{fig:m_ef}. The inset shows the
$f$-magnetization. All three methods (SDA, MAA and MPT) do have a
self-consistent ferromagnetic solution for $e_{\rm f}$ well below the
conduction band. An inspection of the internal energies $\langle H
\rangle$, which can be calculated analogously to~(\ref{eq:nf_ns})
and~(\ref{eq:bandshift2}), of the
ferromagnetic and the paramagnetic solution reveals, however,
that for the SDA and MAA the ferromagnetic solutions are not the stable
ones, the system is in fact paramagnetic. This is not the case for the
MPT.
Here the ferromagnetic solution 
remains stable. This indicates that in the Kondo regime, the low-energy
properties become much more important concering ferromagnetic
ordering. This was first proposed in reference~\cite{TJF97}, where the
origin of ferromagnetic order in the Kondo regime was identified as
RKKY-like. The polarization of the conduction band is due to the
formation of Kondo screening clouds\cite{TJF97}. As was already argued
in section~\ref{sec:approx}, the MAA and the SDA are not able to
reproduce such a mechanism whereas the MPT should contain this at least
qualitatively.

So whereas in the intermediate-valence regime, the SDA, the MAA and the
MPT show similar results, they give completely different pictures
in the Kondo regime. This leads us to the conclusion that there have to
be two distinct mechanisms driving the ferromagnetic ordering in these
two different areas in parameter space. Whereas in the Kondo regime, a
RRKY mechanism is doubtless the key factor, as discussed above and in
reference~\cite{TJF97}, the situation is clearly different in the
intermediate-valence regime. Here, the correct reproduction of
high-energy features, the charge excitation as ensured by the bandshift
correction~(\ref{eq:bandshift}) seems crucial. The inclusion of the
particular low-energy properties of the PAM does not significantly change
the behaviour. 
From our observations, we are led to propose a single-band mechanism
similar to the one leading to ferromagnetism in the single-band Hubbard
model to be responsible for the ferromagnetic ordering in the
intermediate-valence regime.

This proposal is further supported by the following
observations:\\
I) The critical interaction strength
$U_{\rm c}$ is much larger in the intermediate-valence regime than in
the Kondo
regime thus pointing to a genuine strong-coupling effect (cf.\
reference~\cite{mn00c}).\\
II) The lower critical $n^{\text{(tot)}}$ marking the breakdown of
ferromagnetism is in fact determined by a critical
$n^{(f)}$\cite{mn00c}. The conduction electron density has no
influence on the magnetic phase boundaries.
This is a clear reference to a single-band mechanism.\\
III) The polarization of the conduction band shows a remarkable
behaviour as e.g.\ its change of sign. This, however,
does not affect important magnetic quantities as the Curie
temperature. The polarization of the
conduction electrons seems to be a consequence and not the cause of the
ferromagnetic ordering of the $f$ electrons.

From these points, we arrive at the proposition that the ferromagnetic
order in the intermediate-valence regime is due to some intra-band
mechanism. For the Hubbard model\cite{Hub63,Gut63,Kan63}, the existence
of a ferromagnetic phase in the strong-coupling regime was confirmed in
the limit of infinite dimensions\cite{Ulm98,OPK97}. The mechanism
driving this transition is simply based on a gain of kinetic
energy\cite{Vea97,PHWN98}. This is supported by the strong dependency on the
shape of the free ($U=0$) density of states\cite{HN97b,Wea98}. It was
confirmed that ferromagnetism is most favored in case of a non-symmetric
DOS which has a divergence at or close to one of its edges. 
Going back to the PAM, we note that the hybridization leads in the
intermediate-valence regime to an
effective $f$-$f$-hopping. The $f$-electrons form a
strongly correlated band. This band fits well into the prerequisits
of a single-band ferromagnet as lined out above: The band is narrow,
strongly asymetric and most of its spectral weight is, for appropriate
values of $e_f$, located near its edge (cf.\ figure~\ref{fig:dos_u0},
thick lines). 
The proposed similarity between ferromagnetism in
the intermediate-valence regime
of the PAM and the Hubbard model manifests itself also in the fact that in both
cases, the fulfillment of the high-energy expansion~(\ref{eq:expansion})
(as done by SDA, MAA and MPT) seems crucial for a
proper description of the phenomenon\cite{PHWN98}.

\section{Summary}
\label{sec:sum}
In this paper, we have discussed ferromagnetism in the periodic Anderson
model (PAM), and possible mechanisms driving the magnetic ordering.

We have reviewed a series of approximation schemes, from the Hubbard-I,
via the spectral-density approximation (SDA) and the modified alloy
analogy (MAA) to
the modified perturbation theory (MPT). This series represents a subsequent
improvement according to several exactly known properties of the
model. The Hubbard-I approximation is exact in the two limiting cases of
vanishing hybridization ($V=0$) and interaction ($U=0$). Its systematic
improvement with respect to the correct reproduction of the high-energy
expansion of the self-energy leads directly to the spectral density
approximation. The inclusion of quasiparticle damping effects without
loosing the correct high-energy behaviour is possible via the modfied
alloy analogy procedure. Finally, the modified perturbation theory still
recovers the correct high-energy expansion, includes quasiparticle
damping effects, and additionally incorporates, at least qualitatively
correct, the special low-energy properties of the PAM.

The results of the SDA, MAA and MPT compare well in the
intermediate-valence regime. By comparing with the Hubbard-I
approximation, it becomes clear that the correct reproduction of the
high-energy behaviour is crucial for a correct description of
ferromagnetism in this parameter regime. The influence of quasiparticle
damping is, as expected, a reduction of the magnetic stability as
indicated by a strongly reduced Curie temperature and smaller
ferromagnetic area in the $T=0$ phase diagram. The low-energy physics
seem to have only minor effects on the ferromagnetic properties.

In the Kondo regime, the picture is completely different. Only the
MPT yields a stable ferromagnetic phase. This, however, is in agreement
with QMC calculations\cite{TJF97}. The SDA and MAA fail to recover these
results. The origin of the ferromagnetic ordering in the Kondo regime is
an RKKY exchange as discussed in reference~\cite{TJF97}. We showed in
this paper why the SDA and MAA cannot reproduce such a mechanism.

However, the good qualitative agreement between the SDA, MAA and MPT
results in the intermediate-valence regime let us believe that here the
driving mechanism towards the ferromagnetic transition must be of
different nature. Our results gave some hints that this mechanism is
similar to that driving the ferromagnetic ordering in the single-band
Hubbard model with a band formed by the effectively
delocalized $f$ electrons.

\begin{acknowledgement}
We acknowledge fruitful discussions with our collegues A. Ramakanth and
G. G. Reddy from Kakatiya University, Warangal (India), in the process
of adapting the MAA to the periodic Anderson model. This collaboration was
financially supported by the \textit{Volkswagen foundation}. One of the authors
(D.\  M.\ ) further acknowledges support from the \textit{Friedrich-Naumann
foundation}. 
\end{acknowledgement}

\end{document}